\input harvmac

%%%
%%%
%%% this is the revised version
%%%
%%%

\def\CJ{{\cal J}}

\def\ap{\alpha'}
\def\si{{\sigma^1}}
\def\sii{{\sigma^2}}

\def\Ga{{\Gamma}}

\def\K3{{\bf K3}}
\def\journal#1&#2(#3){\unskip, \sl #1\ \bf #2 \rm(19#3) }
\def\andjournal#1&#2(#3){\sl #1~\bf #2 \rm (19#3) }

\def\bar{\overline}
\def\hat{\widehat}

\def\tilde{\widetilde}

\def\frac#1#2{{#1\over#2}}

\def\half{\frac12}

\def\inbar{\,\vrule height1.5ex width.4pt depth0pt}
\def\IC{\relax\hbox{$\inbar\kern-.3em{\rm C}$}}
\def\IR{\relax{\rm I\kern-.18em R}}
\def\IP{\relax{\rm I\kern-.18em P}}

%
%%%%%%%%%%%%%%%%%%%%%%%%%%%%%%%%%%%%
%

%\def\ap#1#2#3{Ann. Phys. {\bf #1} (#2) #3}

%
\catcode`\@=11
\def\slash#1{\mathord{\mathpalette\c@ncel{#1}}}
\overfullrule=0pt

\def\CC{{\cal C}}

\def\underrel#1\over#2{\mathrel{\mathop{\kern\z@#1}\limits_{#2}}}

\catcode`\@=12

%%%%%%%%%%%%%%%%%%%%%%%%%%%%%%%%%%%%%%%%%%%%%%%%%%%%%%%%%%%%%%
\def\sdtimes{\mathbin{\hbox{\hskip2pt\vrule height 4.1pt depth -.3pt width
.25pt \hskip-2pt$\times$}}}

\def\exp{{\rm exp}}

%%%%%%%%%%%%%%%%%%%%%%%%%%%%%%%%%%%%%%%%%%%%%%%%%%%%%%%%%%%%%%
% new defs:

\def \ov {\over}
\def \p {\partial}
\def \ha {{1 \ov 2}}
\def \al {\alpha}

\def \ep {\epsilon}

\def \apr {\alpha'}
\def \m {{\rm m}}

\def\IL{\relax{\rm I\kern-.18em L}}
\def\IH{\relax{\rm I\kern-.18em H}}
\def\IR{\relax{\rm I\kern-.18em R}}
\def\IC{\relax\hbox{$\inbar\kern-.3em{\rm C}$}}
\def\IZ{{\bf Z}}
\def\CP {{\cal P }}

\def\CH{{\cal H}}
\def\CU{{\cal U}}

%\def\IZ{{\bf Z}}
%\def\IR{{\bf R}}

%%%%%%%%%%%%%%%%%%%%%%%%%%%%

%%%% you need these macros:
%%%%%
%%%%%

%%% MACROS FOR BOX BOUNDARY CONDS
%%% FROM KAWAI ET AL

\def\makeblankbox#1#2{\hbox{\lower\dp0\vbox{\hidehrule{#1}{#2}%
   \kern -#1% overlap rules
   \hbox to \wd0{\hidevrule{#1}{#2}%
      \raise\ht0\vbox to #1{}% vrule height
      \lower\dp0\vtop to #1{}% vrule depth
      \hfil\hidevrule{#2}{#1}}%
   \kern-#1\hidehrule{#2}{#1}}}%
}%
\def\hidehrule#1#2{\kern-#1\hrule height#1 depth#2 \kern-#2}%
\def\hidevrule#1#2{\kern-#1{\dimen0=#1\advance\dimen0 by #2\vrule
    width\dimen0}\kern-#2}%
\def\openbox{\ht0=1.2mm \dp0=1.2mm \wd0=2.4mm  \raise 2.75pt
\makeblankbox {.25pt} {.25pt}  }

\def\bun#1/#2{\leavevmode
   \kern.1em \raise .5ex \hbox{\the\scriptfont0 #1}%
   \kern-.1em $/$%
   \kern-.15em \lower .25ex \hbox{\the\scriptfont0 #2}%
}

\def\opensquare{\ht0=3.4mm \dp0=3.4mm \wd0=6.8mm  \raise 2.7pt
\makeblankbox {.25pt} {.25pt}  }

%%%%%%%%%%%%%%%%%%%%%%%

\def\sector#1#2{\ {\scriptstyle #1}\hskip 1mm
\mathop{\opensquare}\limits_{\lower 1mm\hbox{$\scriptstyle#2$}}\hskip 1mm}

\def\tsector#1#2{\ {\scriptstyle #1}\hskip 1mm
\mathop{\opensquare}\limits_{\lower 1mm\hbox{$\scriptstyle#2$}}^\sim\hskip 1mm}
%%%
%%%

\def\CP{{\cal P}}

\def\ap{\alpha'}

\def\IZ{{\bf Z}}

%%%%%%%%%%%%%%%%%%%%%%%%%%%%%%%%%%%%%%%%%%%%%%%%%%%

\lref\polchinski{J. Polchinski, ``Superstring Theory'', Vol. 1,
Cambridge University Press, Cambridge, 1998.}

\lref\hkmm{J.A. Harvey, D. Kutasov, E. Martinec, and G. Moore,
``Localized Tachyons and RG Flows,''  hep-th/0111154 }

\lref\deconstruction{N. Arkami-Hamed, A.G. Cohen, D.B. Kaplan, A.
Karch, and L. Motl, ``Deconstructing $(2,0)$ and Little String
Theories'' hep-th/0110146}

\lref\atickwitten{J. Atick and E. Witten, ``The Hagedorn
transition and the number of degrees of freedom of string
theory,'' Nucl.Phys.B310:291,1988 }

\lref\egkr{S.~Elitzur, A.~Giveon, D.~Kutasov and E.~Rabinovici, to
appear}

\lref\rohm{R. Rohm, ``Spontaneous supersymmetry breaking in
supersymmetric string theories,''  Nucl.\ Phys.\ B {\bf 237}, 553 (1984). }

\lref\finiteall{G. Moore, ``Finite in All Directions,''
hep-th/9305139}

%\KoganNN
\lref\KoganNN{ I.~I.~Kogan and N.~B.~Reis, ``H-branes and chiral
strings,'' Int.\ J.\ Mod.\ Phys.\ A {\bf 16}, 4567 (2001)
[arXiv:hep-th/0107163].
%%CITATION = HEP-TH 0107163;%%
}

\lref\horsteif{G.T. Horowitz and A.R. Steif, ``Is spacetime
duality violated in time dependent string solutions?'' Phys.
Lett. {\bf 250B}(1990)49}

\lref\smithpolchinski{E. Smith and J.
Polchinski, ``Duality survives time dependence,'' Phys. Lett.
{\bf B} (1991) 59}

%\HorowitzAP
\lref\HorowitzAP{ G.~T.~Horowitz and A.~R.~Steif, ``Singular
String Solutions With Nonsingular Initial Data,'' Phys.\ Lett.\ B
{\bf 258}, 91 (1991).
%%CITATION = PHLTA,B258,91;%%
}

%\SusskindIF
\lref\SusskindIF{ L.~Susskind, L.~Thorlacius and J.~Uglum, ``The
Stretched horizon and black hole complementarity,'' Phys.\ Rev.\
D {\bf 48}, 3743 (1993) [arXiv:hep-th/9306069].
%%CITATION = HEP-TH 9306069;%%
}

%\BirrellIX
\lref\BirrellIX{ N.~D.~Birrell and P.~C.~Davies, ``Quantum Fields
In Curved Space,'' {\it  Cambridge, Uk: Univ. Pr. ( 1982) 340p}. }

%\VenezianoPZ
\lref\VenezianoPZ{ G.~Veneziano, ``String cosmology: The pre-big
bang scenario,'' arXiv:hep-th/0002094.
%%CITATION = HEP-TH 0002094;%%
}

%\KhouryBZ
\lref\KhouryBZ{ J.~Khoury, B.~A.~Ovrut, N.~Seiberg,
P.~J.~Steinhardt and N.~Turok, ``From big crunch to big bang,''
arXiv:hep-th/0108187;
%%CITATION = HEP-TH 0108187;%%
%\SeibergHR
N.~Seiberg, ``From big crunch to big bang - is it possible?,''
arXiv:hep-th/0201039.
%%CITATION = HEP-TH 0201039;%%
}

\lref\BanadosWN{ M.~Banados, C.~Teitelboim and J.~Zanelli, ``The
Black Hole In Three-Dimensional Space-Time,'' Phys.\ Rev.\ Lett.\
{\bf 69}, 1849 (1992) [arXiv:hep-th/9204099].
%%CITATION = HEP-TH 9204099;%%
}
%\BanadosGQ
\lref\BanadosGQ{ M.~Banados, M.~Henneaux, C.~Teitelboim and
J.~Zanelli, ``Geometry of the (2+1) black hole,'' Phys.\ Rev.\ D
{\bf 48}, 1506 (1993) [arXiv:gr-qc/9302012].
%%CITATION = GR-QC 9302012;%%
}
%\ShiraishiHF
\lref\ShiraishiHF{ K.~Shiraishi and T.~Maki, ``Quantum fluctuation
of stress tensor and black holes in three-dimensions,'' Phys.\
Rev.\ D {\bf 49}, 5286 (1994).
%%CITATION = PHRVA,D49,5286;%%
}

%\SteifZV
\lref\SteifZV{ A.~R.~Steif, ``The Quantum Stress Tensor In The
Three-Dimensional Black Hole,'' Phys.\ Rev.\ D {\bf 49}, 585
(1994) [arXiv:gr-qc/9308032].
%%CITATION = GR-QC 9308032;%%
}

%\LifschytzEB
\lref\LifschytzEB{ G.~Lifschytz and M.~Ortiz, ``Scalar Field
Quantization On The (2+1)-Dimensional Black Hole Background,''
Phys.\ Rev.\ D {\bf 49}, 1929 (1994) [arXiv:gr-qc/9310008].
%%CITATION = GR-QC 9310008;%%
}

%\PolchinskiMF
\lref\PolchinskiMF{ J.~Polchinski, ``Critical Behavior Of Random
Surfaces In One-Dimension,'' Nucl.\ Phys.\ B {\bf 346}, 253
(1990).
%%CITATION = NUPHA,B346,253;%%
}

%\PolchinskiUQ
\lref\PolchinskiUQ{ J.~Polchinski, ``Classical Limit Of
(1+1)-Dimensional String Theory,'' Nucl.\ Phys.\ B {\bf 362}, 125
(1991).
%%CITATION = NUPHA,B362,125;%%
}

%\DiFrancescoSS
\lref\DiFrancescoSS{ P.~Di Francesco and D.~Kutasov,
``Correlation functions in 2-D string theory,'' Phys.\ Lett.\ B
{\bf 261}, 385 (1991).
%%CITATION = PHLTA,B261,385;%%
}

%\DiFrancescoUD
\lref\DiFrancescoUD{ P.~Di Francesco and D.~Kutasov, ``World
sheet and space-time physics in two-dimensional (Super)string
theory,'' Nucl.\ Phys.\ B {\bf 375}, 119 (1992)
[arXiv:hep-th/9109005].
%%CITATION = HEP-TH 9109005;%%
}

%\MooreGB
\lref\MooreGB{ G.~W.~Moore and R.~Plesser, ``Classical scattering
in (1+1)-dimensional string theory,'' Phys.\ Rev.\ D {\bf 46},
1730 (1992) [arXiv:hep-th/9203060].
%%CITATION = HEP-TH 9203060;%%
}

%\KimMI
\lref\KimMI{ H.~Kim, J.~S.~Oh and C.~R.~Ahn, ``Quantisation of
conformal fields in AdS(3) black hole spacetime,'' Int.\ J.\ Mod.\
Phys.\ A {\bf 14}, 2431 (1999) [arXiv:hep-th/9708072].
%%CITATION = HEP-TH 9708072;%%
}
%\CarlipQV
\lref\CarlipQV{ S.~Carlip, ``The (2+1)-Dimensional black hole,''
Class.\ Quant.\ Grav.\  {\bf 12}, 2853 (1995)
[arXiv:gr-qc/9506079].
%%CITATION = GR-QC 9506079;%%
}
%\MaldacenaKR
\lref\MaldacenaKR{ J.~M.~Maldacena, ``Eternal black holes in
Anti-de-Sitter,'' arXiv:hep-th/0106112.
%%CITATION = HEP-TH 0106112;%%
} \lref\HorowitzAP{ G.~T.~Horowitz and A.~R.~Steif, ``Singular
String Solutions With Nonsingular Initial Data,'' Phys.\ Lett.\ B
{\bf 258}, 91 (1991).
%%CITATION = PHLTA,B258,91;%%
}
%\BalasubramanianRY
\lref\BalasubramanianRY{ V.~Balasubramanian, S.~F.~Hassan,
E.~Keski-Vakkuri and A.~Naqvi, ``A space-time orbifold: A toy
model for a cosmological singularity,'' arXiv:hep-th/0202187.
%%CITATION = HEP-TH 0202187;%%
}
%\GutperleAI
\lref\GutperleAI{ M.~Gutperle and A.~Strominger, ``Spacelike
branes,'' arXiv:hep-th/0202210.
%%CITATION = HEP-TH 0202210;%%
}
%\CornalbaFI
\lref\CornalbaFI{ L.~Cornalba and M.~S.~Costa, ``A New
Cosmological Scenario in String Theory,'' arXiv:hep-th/0203031.
%%CITATION = HEP-TH 0203031;%%
}
%\NekrasovKF
\lref\NekrasovKF{ N.~A.~Nekrasov, ``Milne universe, tachyons, and
quantum group,'' hep-th/0203112.
%%CITATION = HEP-TH 0203112;%%
}

%\SenNU
\lref\SenNU{
A.~Sen,
``Rolling Tachyon,''
arXiv:hep-th/0203211.
%%CITATION = HEP-TH 0203211;%%
}

%\HiscockJQ
\lref\HiscockJQ{ W.~A.~Hiscock, ``Quantized fields and chronology
protection,'' arXiv:gr-qc/0009061.
%%CITATION = GR-QC 0009061;%%
}

%\HawkingNK
\lref\HawkingNK{ S.~W.~Hawking, ``The Chronology protection
conjecture,'' Phys.\ Rev.\ D {\bf 46}, 603 (1992).
%%CITATION = PHRVA,D46,603;%%
}

%\HawkingPK
\lref\HawkingPK{ S.~W.~Hawking, ``The Chronology Protection
Conjecture,'' {\it Prepared for 6th Marcel Grossmann Meeting on
General Relativity (MG6), Kyoto, Japan, 23-29 Jun 1991}. }

%\BrodskyDE
\lref\BrodskyDE{ S.~J.~Brodsky, H.~C.~Pauli and S.~S.~Pinsky,
``Quantum chromodynamics and other field theories on the light
cone,'' Phys.\ Rept.\  {\bf 301}, 299 (1998)
[arXiv:hep-ph/9705477].
%%CITATION = HEP-PH 9705477;%%
}

%\SimonMA
\lref\SimonMA{ J.~Simon, ``The geometry of null rotation
identifications,'' arXiv:hep-th/0203201.
%%CITATION = HEP-TH 0203201;%%
}

\lref\mwaver{JM.~Figueroa-O'Farrill, ``Breaking the M-waves,''
Class.\ Quant.\ Grav.\ {\bf 17}, 2925 (2000)
[arXiv:hep-th/9904124].}

\lref\longpaper{H. Liu, G. Moore, and N. Seiberg, To appear.}

\lref\polchinski{J. Polchinski, ``Superstring Theory'', Vol. 1,
Cambridge University Press, Cambridge, 1998.}

\lref\BanadosWN{ M.~Banados, C.~Teitelboim and J.~Zanelli, ``The
Black Hole In Three-Dimensional Space-Time,'' Phys.\ Rev.\ Lett.\
{\bf 69}, 1849 (1992) [arXiv:hep-th/9204099].
%%CITATION = HEP-TH 9204099;%%
}
%\BanadosGQ
\lref\BanadosGQ{ M.~Banados, M.~Henneaux, C.~Teitelboim and
J.~Zanelli, ``Geometry of the (2+1) black hole,'' Phys.\ Rev.\ D
{\bf 48}, 1506 (1993) [arXiv:gr-qc/9302012].
%%CITATION = GR-QC 9302012;%%
}

\lref\hkmm{J.A. Harvey, D. Kutasov, E. Martinec, and G. Moore,
``Localized Tachyons and RG Flows,''  hep-th/0111154 }

\lref\deconstruction{N. Arkami-Hamed, A.G. Cohen, D.B. Kaplan, A. Karch,
and L. Motl, ``Deconstructing $(2,0)$ and Little String Theories''
hep-th/0110146}

\lref\atickwitten{J. Atick and E. Witten, ``The Hagedorn
transition and the number of degrees of freedom of
string theory,'' Nucl.Phys.B310:291-334,1988 }

%\MooreZC
\lref\MooreZC{
G.~W.~Moore,
``Finite In All Directions,''
arXiv:hep-th/9305139.
%%CITATION = HEP-TH 9305139;%%
}

\lref\horsteif{G.T. Horowitz and A.R. Steif, ``Is spacetime
duality violated in time dependent string solutions?''
Phys. Lett. {\bf 250B}(1990)49}

\lref\smithpolchinski{E. Smith and J. Polchinski,
``Duality survives time dependence,'' Phys. Lett. B {\bf 263} 59 (1991).}

%\HorowitzAP
\lref\HorowitzAP{ G.~T.~Horowitz and A.~R.~Steif, ``Singular
String Solutions With Nonsingular Initial Data,'' Phys.\ Lett.\ B
{\bf 258}, 91 (1991).
%%CITATION = PHLTA,B258,91;%%
}

%\Arkani-HamedIE
\lref\ArkaniHamedIE{
N.~Arkani-Hamed, A.~G.~Cohen, D.~B.~Kaplan, A.~Karch and L.~Motl,
``Deconstructing (2,0) and little string theories,''
arXiv:hep-th/0110146.
%%CITATION = HEP-TH 0110146;%%
}

\lref\tseytlinup{
A.~A.~Tseytlin,
``Exact string solutions and duality,''
arXiv:hep-th/9407099;
%%CITATION = HEP-TH 9407099;%% \cr
C.~Klimcik and A.~A.~Tseytlin, unpublished (1994);
A. A. Tseytlin, unpublished (2001).}

\lref\aharony{
O.~Aharony, M.~Fabinger, G.~Horowitz, and E.~Silverstein,
``Clean Time-Dependent String Backgrounds from Bubble Baths,''
arXiv:hep-th/0204158.}

%\TolleyCV
\lref\TolleyCV{ A.~J.~Tolley and N.~Turok, ``Quantum fields in a
big crunch / big bang spacetime,'' arXiv:hep-th/0204091.
%%CITATION = HEP-TH 0204091;%%
}

\lref\hp{G.T.~Horowitz and J.~Polchinski, to appear.}

\lref\lmst{H.~Liu, G.~Moore and N.~Seiberg, to appear.}

\lref\KiritsisKZ{ E.~Kiritsis and B.~Pioline, ``Strings in
homogeneous gravitational waves and null holography,''
arXiv:hep-th/0204004.
%%CITATION = HEP-TH 0204004;%%
}

\lref\lawrence{A.~Lawrence, ``On the instability of 3d null
singularities'', hep-th/0205288.}

%%%%%%%%%%%%%%%%%%%%%%%%%%%%%%%%%%%%%%%%%%%%%%%%
%\special{"%
%  gsave %
%  %Resolution 72 div dup neg scale %
%  %currentpoint translate % 50 rotate
%  gsave 80 -580 translate 50 rotate %
%  /Times-Roman findfont 180 scalefont setfont %
%  10 -1 0 { dup .03 mul .7 add setgray dup neg moveto (DRAFT) show } for %
%  .9 setgray 0 0 moveto (DRAFT) show %
% % 150 150 translate %
% % /Times-Roman findfont 80 scalefont setfont %
% % .95 setgray 2 setlinewidth 0 0 moveto (\draftbeer) false charpath stroke %
%  grestore} %
%%%%%%%%%%%%%%%%%%%%%%%%%%%%%%%%%%%%%%%%%%%%%%%%%%%%%%%%%%%%%%%

\Title{\vbox{\baselineskip12pt \hbox{hep-th/0204168}
\hbox{RUNHETC-2002-11}
}}%
{\vbox{\centerline{Strings in a Time-Dependent Orbifold} }}

\smallskip
\centerline{Hong Liu, Gregory Moore}
\medskip

\centerline{\it Department of Physics, Rutgers University}
\centerline{\it Piscataway, New Jersey, 08855-0849}

\bigskip

\centerline{and}

\bigskip

\centerline{Nathan Seiberg}
\medskip
\centerline{\it School of Natural Sciences}
\centerline{\it Institute for Advanced Study}
\centerline{\it Einstein Drive,Princeton, NJ 08540}

\smallskip

\vglue .3cm

\bigskip
\noindent We consider string theory in a time dependent orbifold
with a null singularity. The singularity separates a contracting
universe from an expanding universe, thus constituting a big
crunch followed by a big bang.  We quantize the theory both in
light-cone gauge and covariantly.  We also compute some tree and
one loop amplitudes which exhibit interesting behavior near the
singularity.  Our results are compatible with the possibility
that strings can pass through the singularity from the
contracting to the expanding universe, but they also indicate the
need for further study of certain divergent scattering amplitudes.

\Date{Revised: June 7, 2002}

%\draftmode

%%%%%%%%%%%%%%%%%%%%%%%%%%%%%%%%%

\newsec{Introduction and Motivation}

Much of the work in string theory and CFT has been in the context
of {\it Euclidean signature} target space, i.e. in {\it
time-independent } background geometries. Many interesting
problems in physics involve time in an essential way. It is
therefore of interest to extend the study of string theory to
time-dependent backgrounds.

The extension of string theory from Euclidean to Lorentzian
signature is nontrivial and many new issues arise. For example,
in a time-dependent background there is no natural definition of
the vacuum.  Also, it is not always clear what the correct
observables of string theory are.  Another question we may ask is
whether time comes to an end. If it does, how do we describe the
boundary conditions or final states. If it does not, how do
strings resolve or pass through the spacelike singularities
predicted by general relativity.  Such singularities arise behind
the horizon of black holes and in the big-bang.  Therefore
understanding them is of great  interest. We may also ask whether
timelike and null closed curves are pathological, and if not,
whether there are interesting string-winding effects associated
with such curves. The list of questions and issues goes on, but
the above questions are representative.

In this paper we begin an exploration of these questions in
string theory by describing a simple model of a time-dependent
geometry in which one can attempt to address some of the above
issues in a controlled setting.  In a companion longer paper
\longpaper\ we will provide more details.

In section 2 we describe the model and its geometry.  In section
3 we study the functions on our spacetime which are the wave
functions of the first quantized particles.  In sections 4 and 5
we quantize free strings in the light-cone and conformal gauges
and compute the torus partition function.  Section 6 is devoted
to a preliminary analysis of the interactions and backreaction.
Our conclusions are presented in section 7.

%%%%%%%%%%%%%%%%%%%%%%%%%%%%%%%%%%%%
\newsec{Geometry of the Orbifold}

%%%%%%%%%%%%%%%%%%%%%%%%%%%%%%%%%%%%%%%

\subsec{The model}

Time dependent backgrounds are difficult to work with in general.
As with Calabi-Yau compactification, orbifolds provide a useful
approach: They are simple enough to be solvable, yet complicated
enough to illustrate nontrivial effects. An interesting class of
time-dependent models is based on target spaces of the form
$\bigl( \IR^{1,n}/\Gamma \bigr) \times \CC^{\perp}$ where the
orbifold group is a discrete subgroup $ \Gamma$ of the Poincar\'e
group, and $\CC^\perp$ is a ``transverse'' conformal field theory
rendering the full string theory consistent\foot{One can also
consider the case where $\Gamma$ also acts on ``transverse''
coordinates.}. This class of models was discussed about 12 years
ago by Horowitz and Steif \HorowitzAP.  Recently there has been a
renaissance in the subject motivated in part by the desire to use
string theory to address questions of cosmology
\refs{\KhouryBZ\BalasubramanianRY\CornalbaFI\NekrasovKF
\SimonMA-\TolleyCV}.  Other time dependent backgrounds were
studied in
\refs{\KoganNN\GutperleAI\SenNU\aharony\KiritsisKZ-\egkr}.

The model studied in this letter is based on the target space
$\bigl(\IR^{1,2}/\Gamma \bigr) \times \CC^{\perp}$ where
$\Gamma\cong \IZ$ is a subgroup of the 3D Lorentz group ${\rm
Spin}(1,2) \cong SL(2,R)$. The orbifold group  $\Gamma$ is
completely specified by choosing a conjugacy class of a generator
$g_0$. $SL(2,R)$ has three distinct conjugacy classes,  elliptic,
hyperbolic, and parabolic. The elliptic classes correspond to
spatial rotations, the hyperbolic classes correspond to boosts
leaving one spatial dimension fixed, and the parabolic classes
correspond to ``null boosts.'' We will choose $g_0$ to be a
parabolic element.  This leads to the ``null orbifold''
introduced in \HorowitzAP, briefly studied in \tseytlinup\ and
more recently considered in \SimonMA. We will now describe the
geometry of this orbifold in some detail.

We introduce coordinates $x^\mu$ on  $\IR^{1,2}$ which are
assembled into a column vector $X$. The Lorentz metric is
$ds^2=-2dx^+dx^-+dx^2$. The generator $g_0$ acts as
 \eqn\partrans{ X:=
 \pmatrix{ x^+ \cr x^{~}\cr x^-\cr} \quad \rightarrow\quad
 g_0\cdot X=e^{v\CJ} X = \pmatrix{ x^+ \cr x + v x^+  \cr
 x^- + v x + \half v^2 x^+ \cr} ;\qquad \CJ=\pmatrix{0&0&0\cr
 1&0&0\cr0&1&0}
 }
That is, $ g_0= \exp\bigl(i v J\bigr)$ where we take the Lie
algebra generator
 \eqn\liegen{ J = {1\over \sqrt{2}}(J^{0x} + J^{1x} )}
corresponding to a linear combination of a boost and a rotation.
By a boost in the 1-direction we can set    $v = 2\pi$.

The geodesic distance between a point and its $n$'th image is
$|nvx^+|$.  Therefore our orbifold has no closed timelike
curves.  For $x^+\not=0$ all closed curves are spacelike and for
$x^+=0$ there exist closed null curves.

The orbifold by $\Gamma$ breaks the Poincar\'e symmetry, leaving
only two of its Lie algebra generators unbroken.  These are $J$
of \liegen\ and $p^+=-p_-$ which shifts $x^-$ by a constant.  The
null Killing vector associated with $p^+$ allows us to pick
light-cone gauge; i.e.\ to treat $x^+$ as time.  Since $p^-=-p_+$
is broken by the orbifold the light-cone system depends on $x^+$
and hence it is still nontrivial.  Having a null Killing vector
has an important consequence. The light-cone evolution is first
order in light-cone time and hence it is simpler than standard
second order time evolution.  As a result of that, even though our
background is time dependent, there is no particle production in
the second quantized theory when described in the light-cone
frame.

Equation \partrans\ determines the action of the group element
$g_0$ on spinors up to a sign.  For an appropriate choice of this
sign the group $\Gamma$ leaves one spinor invariant, and therefore
the orbifold has a covariantly constant spinor.  When the
superstring is compactified on this orbifold it preserves half of
the supercharges.  These supercharges square to the Killing vector
$p^+$.

In the light-cone frame where $x^+$ is taken to be the time
$x^+=\tau$ the transformation \partrans\ has the following
physical interpretation.  It is simply a Galilean boost by
velocity $v$. It leaves the time $x^+=\tau$ invariant and shifts
the coordinate $x \to x+ v\tau$. The action of the parabolic
generator $g_0$ on the translation generators $P$ is similar to
\partrans
 \eqn\partransp{P=\pmatrix{p^+\cr p\cr p^-} \to e^{v \CJ} P =
 \pmatrix {p^+\cr p + v p^+ \cr p^-+ v p + \half v ^2p^+}.}
Since $m^2=2p^+p^--p^2$ is Lorentz invariant,  it is invariant
under the parabolic generator $J$.  An analogy to Newtonian
physics emerges when we solve for $p^-={p^2+m^2\over 2p^+}$.  In
the light-cone frame $p^+$ is interpreted as the mass $\mu=p^+$,
$V={m^2\over 2p^+}$ is the potential energy and $p^-={p^2\over
2\mu} +V$ is the total energy.  In terms of the variables
$\mu,p,V$ the parabolic transformation \partransp\ is simply
$p\to p+v\mu$ with $\mu$ and $V$ left invariant.

Thus the parabolic orbifold obtained from \partrans\ can be
considered in the light-cone frame as the quotient by a Galilean
boost.

%%%%%%%%%%%%%%%%%%%%%%%%%%%%%%%%%%%%%%%%%%%%
\subsec{A Little Model for a Big Bang }

It is convenient describe the geometry of the quotient space
$\CO=\IR^{1,2}/\Gamma$ by introducing new coordinates
 \eqn\yspace{ \eqalign{ y^+  & := x^+   \cr y  & :=  {x\over x^+
 }   \cr y^- &  := {2x^+ x^- - x^2\over 2 x^+}= x^- -
 \half{x^2\over x^+} . \cr}}
The advantage of this change of coordinates is that the
identifications are simple (henceforth we take $v=2\pi$):
 \eqn\ycoord{(y^+, y, y^-) \sim (y^+, y+2\pi , y^-) }
and so is the metric
 \eqn\ymetric{\eqalign{
 ds^2 & = -2dx^+ dx^- + (dx)^2 = -2dy^+ dy^- + (y^+)^2 (dy)^2.
 \cr}}
The spacetime \ymetric, which we call the parabolic pinch, may be
visualized as two cones (parametrized by $y^+$ and $y$) with a
common tip at $y^+=0$, crossed with the real line (for $y^-$).
$y$  plays the role of an ``angular variable'' and the null
coordinate $y^+$ plays the role of a ``radial variable.'' As a
function of the ``light-cone time'' $y^+$ we have a big crunch of
the circle at $y^+=0$ which is followed by a big bang. The dual
role of $y^+$ as both a radial variable and a time variable will
be the source of some interesting physics.

It is important that the model does not have closed timelike
curves.  The closed loop parametrized by $y$ is spacelike for
$y^+\not=0$.  At the singularity where $y^+=0$ the circumference
of the $y$ circle vanishes. We will discuss the singularity in
more detail below.

Having in mind light-cone frame, we will refer to all the points
in the orbifold with $x^+=y^+<0$ as the past cone, and to all the
points in the orbifold with $x^+=y^+>0$ as the future cone. An
interesting feature of the spacetime \ymetric, which is
intuitively what we expect from a big crunch followed by a big
bang, is that every point  $\CP = (y^+,y,y^-)$ with $y^+>0$ in the
future cone is in the causal future of every point $\tilde \CP =
(\tilde y^+,\tilde y,\tilde y^-)$ with $\tilde y^+<0$ in the past
cone. This follows since the Lorentzian distance square between
$\CP$ and the $n^{th}$ image $g_0^n \tilde \CP$ of $\tilde \CP$
can be computed from
 \eqn\dissquare{||X-g_0^n \tilde X||^2 =  - 2 \Delta x^+ \Delta
 x^- + (\Delta x)^2 + (2 \pi n)^2 x^+ \tilde x^+ + 2 (2 \pi n)
 (x^+ \tilde x - x \tilde x^+)}
where $\Delta x^{\mu} = x^{\mu} - \tilde x^{\mu}$. At large $n$
the term $(2 \pi n)^2 x^+ \tilde x^+$ dominates, so if $x^+\tilde
x^+<0$, there are infinitely many $g\in \Gamma$ such that $\CP \in
I^+(g\cdot \tilde \CP)$.

We may now formulate our motivating questions more precisely in
this context, namely:
\item{1.}  What is the nature of the singularity at $x^+=y^+=0$?
What happens to string theory there?
\item{2.}  Is the singularity an end of spacetime; i.e.\ does a
consistent formulation of string theory on $\IR^{1,2}/\Gamma$
require {\it one cone} or {\it two cones}? Or is this a choice of
physical model?

%%%%%%%%%%%%%%%%%%%%%%%%%%%%%%%%%%%%%%%%%
\subsec{Nature of the singularity}
%%%%%%%%%%%%%%%%%%%%%%%%%%%%%%%%%%%%%%%%%

In this subsection we analyze the singular subspace $x^+=y^+=0$.
The identification on this subspace is
 \eqn\xpzid{\pmatrix{x^+=0\cr x \cr x^-} \sim  \pmatrix{x^+=0
 \cr x \cr x^- + 2\pi n x}}
Unlike the situation for $x^+\not=0$, here $x$ is not subject to
identification and hence it is a good coordinate.  Note also that
there is no nontrivial identification at $x=0$, so all points
along the $x^-$ axis are distinct. On the other hand, for
very  small $x$, points with very different $x^-$, with
very small spacing,
will be identitfied.  This situation is impossible to describe
using the $y$ coordinates.

Since the coordinate transformation \yspace\ is singular at
$x^+=0$ we need to describe the space $C_Y$ coordinatized by $y$
with some care. \ymetric\ implies that we begin with
$(y^+,y,y^-)\in \IR^3$, and quotient by the equivalence relation
$(y^+,y,y^-)\sim (y^+,y+2\pi ,y^-) $ together with $(y^+=0,y,y^-)
\sim (y^+=0,0,y^-)$. This latter identification is natural since
$y$ is an angular variable and $y^+$ behaves like a radial
variable. More precisely, $C_Y$ projects to 2-dimensional
Minkowski space parametrized by $(y^+,y^-)$ with generic fiber a
circle, except at $y^+=0$, where the fiber degenerates to a
point. $C_Y$ has no closed timelike or null curves and is
Hausdorff.  It looks like a double cone times a line.

In the above we gave a precise definition of the space $C_Y$
following from \ymetric. This space is  in fact not precisely the
orbifold $\CO=\IR^{1,2}/\Gamma$ of section 2,   but is closely
related to it. Our change of coordinates is a continuous map

\eqn\pimap{ \pi: C_Y \to \CO=\IR^{1,2}/\Gamma }
given explicitly by
\eqn\pimapii{ \eqalign{ x^+ & = y^+ \cr x & = y y^+ \cr x^- & =
y^- + \half y^+ y^2 \cr} }
This is an isomorphism for $x^+ = y^+ \not=0$, but is not even
surjective for $y^+=0$. In fact, the space $\CO$ is not Hausdorff.
Recall that the Hausdorff separation axiom states that open sets
separate distinct points.  That is, $\forall P\not=Q, $ $
\exists$ open sets $  U_P,U_Q $ containing $P,Q$, respectively,
such that $  U_P\cap U_Q = \emptyset$. To illustrate the
non-Hausdorff nature of the orbifold consider the simplified
problem of the quotient of the plane $x^+=0$ by the
identification \xpzid. As we have have mentioned, all points
along the $x^-$ axis ($x^+=x=0$) are distinct. On the other hand,
for small $x$ the open sets which resolve different values of
$x^-$ must also get small, and this leads to a non-Hausdorff
topology. Specifically, one finds that points on the lines $L_A =
\{ (0,A,x^-): x^-\in \IR\} $ and $L_{-A} = \{ (0,-A,x^-): x^-\in
\IR\} $ cannot be separated by open sets in $\CO$. Of course,  by
adding an equivalence relation to $\CO$: $(x^+=0, x, x^-) \sim
(x^+=0, x=0, x^-) $ or by considering its subspace which is the
image of \pimap\ we produce a new space $\CO'$ topologically
isomorphic to $C_Y$.

To summarize: in studying our time-dependent string background we
are lead to consider two distinct spaces, the group quotient
$\CO$, which is non-Hausdorff, and the parabolic pinch  $\CO'$,
which is Hausdorff.  They are identical away from the singularity
but the nature of the singularity in the two spaces is
different.  One useful way of thinking about the distinction is
in terms of the foliation by equal-time slices. $\CO$ is foliated
by slices $\CF_{x^+}$, where $\CF_{0}$ is not Hausdorff, while
$\CO'$ is foliated by slices $\CF_{y^+}'$, where $\CF_{0}' =
\{(0,0, y^-) \}$ is a real line. For $x^+=y^+\not=0$ the map
\pimap\ defines an isomorphism of the foliated spaces.

The advantage of the $y$ coordinate system is that it gives us a
clear picture of the topology both of $\CO$ and of $\CO'$ for
$x^+=y^+\not=0$. On $\CO'$, $y^\mu$ is a good global coordinate
system including the singularity on the double cone times a
line. On the group quotient $\CO$ the $y$
coordinate system is singular at $\CF_0$. This is clear from the
identification \xpzid. Therefore, for questions associated with
the singularity of $\CO$ we will prefer to use the $x$ coordinate
system.

The standard string orbifold procedure constructs string theory
on the quotient by a {\it group} action. In formulating strings
in  light-cone gauge it might be possible to construct strings
propagating on $\CO'$, as well as on $\CO$.  Unfortunately, the
consistency of string theory on $\CO'$ is not self-evident since
$\CO'$ is not geodesically complete at $y^+=0$; some geodesics
reach  $y^-\to \pm \infty$ in finite proper time. In the covariant
formulation described below, we are describing strings
propagating on $\CO$.

%%%%%%%%%%%%%%%%%%%%%%%%%%%%%%%%%
\subsec{ Relation to other models}

The parabolic orbifold is closely related to two other models,
which, at first sight, appear to {\it require} that physics only
makes sense on one cone, and not two. We believe that in each of
these examples there is a subtlety which invalidates the argument
for a single cone.

Our first example is the elliptic orbifold: $(C/\IZ_N )\times
\IR$, boosted in the $(r,t)$ plane by a boost $\sim 1/N$, in the
limit as $N\to \infty$. Locally, the metric degenerates to that
of $C_Y$. However, one must carefully consider the {\it range } of
the coordinates. There are regions of spacetime mutually
inaccessible in the initial and final systems: It is not even
true that the one-cone theory is the limit of the elliptic
orbifold.

Our second example is the  $J=M=0$ BTZ black hole
\refs{\BanadosWN,\BanadosGQ}. Let us identify  $\IR^{1,2}$ with
the Lie algebra   of $SL(2,R)$. Then \partrans\ is simply the
adjoint action by $g_0$. Now,   the BTZ black hole is simply
obtained by replacing the Lie algebra by the Lie group
$\widetilde{SL(2,R)}$, \foot{$\widetilde{SL(2,R)}$ is the
universal cover of $SL(2,R)$} and by promoting the adjoint action
of $\Gamma$ on the Lie algebra to the adjoint action on the
group, and restricting to a single fundamental domain. The Lie
algebra is the infinitesimal region of the identity and indeed a
scaling limit of the BTZ metric near the singularity produces
\ymetric. (In this context the fact that the singularity is not
Hausdorff was noted in \BanadosGQ.) One might think that the
AdS/CFT correspondence demands that we do not continue beyond the
singularity. But careful consideration suggests that there is no
immediate contradiction between the AdS/CFT correspondence and the
existence of a possible continuation  beyond the singularity.

%%%%%%%%%%%%%%%%%%%%%%%%%%%%%%%%%%%%%%%
\newsec{First Quantized Theory}
%%%%%%%%%%%%%%%%%%%%%%%%%%%%%%%%%%%%%%%

In this section, as a warmup for string theory on the orbifold,
we consider first quantized particles. The first quantized wave
equation for a  spinless  particle of mass $m$ is:
 \eqn\waveeq{ \bigl[-2 {\p \over \p x^+}{\p \over \p x^-}+ ({\p
 \over \p x})^2 \bigr] \psi = m^2 \psi .}
To define the orbifold Hilbert space we project onto
wavefunctions invariant under
 \eqn\jdifo{\CU(g_0) := \exp(2\pi i \hat J), \qquad
 \hat J = \hat x^+ \hat p - \hat x \hat p^+ = -i \bigl( x^+{\p
 \over \p x} + x {\p \over \p x^-}\bigr) . }
The generators of the Poincare algebra which are invariant under
$\CU (g_0)$ are $\hat p^+$ and $\hat J$. Thus on the orbifold it
is convenient to diagonalize these operators. Explicitly\foot{We
will take $p^+ \not=0$. Some further comments on the $p^+=0$ case
will be found in \longpaper.},
 \eqn\wavefun{\eqalign{ \psi_{p^+,J}  = &\sqrt{p^+\over ix^+}\
 \exp \left[-ip^+x^--i{m^2\over 2p^+} x^+ + i{ p^+
 \over 2 x^+}(x-\xi)^2\right] \cr
& = \int_{-\infty}^\infty {dp  \over \sqrt{2 \pi }}\
 e^{-ip\xi}  \phi_{p^+,p}(x^+,x^-, x)\cr
 }}
where $ \xi := - J/p^+$ and hence the eigenvalue $J$ enters as a
``position'' $\xi$.  In the second line above we have expanded the
wave function in terms of standard on-shell plane wave basis
 \eqn\phidef{\phi_{p^+,p}(x^+,x,
 x^-)=\exp\left(-ip^+x^- -ip^-x^+ +i px\right), \qquad\qquad p^-=
 {p^2+m^2 \over 2p^+}}
The plane waves are not invariant under the orbifold action
\partrans, and the invariant functions obtained by summing over
the images are not a convenient basis on  the orbifold. It is
easy to check that in terms of $\psi_{p^+,J}$ \wavefun\ the
orbifold projection is simply $J \in \IZ$.

Since $\psi_{p^+,J}$ is  a Fourier transformation in $p$ it can
be interpreted as an $x$-eigenfunction. Indeed
 \eqn\psili{\lim_{x^+ \to 0}\psi_{p^+,J} (x^+ ,x, x^-) =
 \sqrt{ 2\pi} e^{-ip^+x^-}\delta(x-\xi),\qquad \qquad \xi=-J/p^+}
This result can be derived more directly by considering an
eigenvector  of $\partial_{x^-}$ which is well defined on the
surface $\CF_0$ under the identification \xpzid.\foot{Note that
the limit \psili\ is not well defined on the Hausdorff space
$\CF_0' \subset \CO'$ since the distribution \psili\ separates
points which are identified in $\CO'$.} Equation \psili\ can also
be understood from the fact that $\hat J = \hat x^+ \hat p- \hat x
\hat p^+ \to - \hat x \hat p^+ $ at $x^+ = 0$. Thus the
eigenfunction of $\hat J$ and $\hat p^+$ must be a coordinate
eigenfunction with eigenvalue $\xi= - J/p^+.$ Since $J$ is
quantized on the orbifold, we see that with fixed $p^+$ the
wavefunctions are supported on the lattice $x\in {1\over p^+}
\IZ$ at $x^+ =0$.

Equation \psili\ should be interpreted with care. The limit of
$\psi$ is a distribution and should only be convoluted with
smooth functions. For example, it is clear that $\lim_{x^+ \to 0}
|\psi_{p^+,J}|^2=\left|{p^+\over x^+} \right| $ which is not
localized at $\xi$.

%%%%%%%%%%%%%%%%%%%%%%%%%%%%%%%%%%%%%%%%%%%%%%%%%%%%
\newsec{ Strings on the orbifold: Light-Cone Gauge }
%%%%%%%%%%%%%%%%%%%%%%%%%%%%%%%%%%%%%%%%%%%%%%%%%%%%

In this section we analyze the system in the light-cone gauge
$x^+=y^+ = \tau$.  The light-cone gauge Lagrangian is
 \eqn\lcgiii{\eqalign{
 L=& - p^+ \partial_\tau x^-_0 + {1\over 4\pi \alpha'}
 \int_0^{2\pi}
 d\sigma \left(\alpha' p^+ \p_\tau x \p_\tau x
 - {1\over \alpha'p^+ } \p_\sigma x \p_\sigma x \right)\cr
 & = -p^+ \partial_\tau y^-_0 + {1\over 4\pi \alpha'}
 \int_0^{2\pi}
 d\sigma \tau^2\left(\alpha' p^+ \p_\tau y\p_\tau y
 - {1\over \alpha'p^+} \p_\sigma y \p_\sigma y \right)}}
where
 \eqn\xyrelli{\eqalign{
 &x(\sigma,\tau)=\tau y(\sigma,\tau) \cr
 &y^-_0(\tau)=x^-_0(\tau)-{1 \over 2\tau} \int_0^{2\pi}
  {d\sigma \over 2\pi}  (x(\sigma, \tau))^2}}
Invariance under constant shifts of $\sigma$ is implemented by
imposing \longpaper
 \eqn\lzmlzb{\int d\sigma \left(\partial_\sigma x \partial_\tau
 x - {1\over 2\tau} \partial_\sigma x^2 \right)= \int d\sigma
 \tau^2 \partial_\sigma y \partial_\tau y=0}
It is important that the two expressions for the Lagrangian
\lcgiii\ and the expressions for the constraint \lzmlzb\ are
invariant under the orbifold identification
 \eqn\identlcg{\eqalign{
 &x(\sigma,\tau) \to x(\sigma,\tau) +2\pi n \tau\cr
 &x^-_0 (\tau) \to x^-_0(\tau) + 2\pi n \int_0^{2\pi}
 {d \sigma \over 2\pi}
 x(\sigma,\tau) +{(2\pi n)^2\over 2} \tau \cr
 &y(\sigma,\tau)\to y(\sigma,\tau) + 2\pi n}}
The equations of motions for $x^-_0$ and  $y^-_0$ set $p^+$ to a
constant.  The equations of motion for $p^+$ leads to
 \eqn\lcgii{\eqalign{
 &P_{x^-}=p^+\partial_\tau x^-_0 = {1\over 4\pi\alpha'}
 \int_0^{\ell} d\sigma \left( \p_\tau x \p_\tau x
 +\p_\sigma x \p_\sigma x \right)\cr
 &P_{y^-}=p^+\partial_\tau y^-_0 ={1\over 4\pi\alpha'}
  \int_0^{\ell} d\sigma \tau^2 \left(\p_\tau y\p_\tau y
 + \p_\sigma y \p_\sigma y \right)}}
where we have rescaled $\sigma$ to range in $[0,\ell=2\pi
\alpha'p^+)$ (we will continue to use this rescaled value).  The
Hamiltonian $P_{x^-}$ is not invariant under the orbifold
identification \identlcg\ but $P_{y^-}$ is invariant. Either
expression can be used in the quantization \longpaper.

A complete set of solutions to the equations of motion in the
$w$-twisted sector, can be expressed in terms of harmonic
oscillators:
 \eqn\xysol{\eqalign{ &
 x(\sigma,\tau) =\xi + {p \over p^+}\tau + {2\pi w
  \sigma \tau\over \ell}+ \cr
 & \qquad i \left({\alpha'\over 2}\right)^\half \sum_{n\not= 0}
 \left\{ {\alpha_n \over n} \exp \left[-
 {2\pi in(\sigma+\tau) \over \ell}\right] + {\tilde \alpha_n \over
 n} \exp\left[ {2\pi in(\sigma-\tau) \over \ell}\right]
 \right\}\cr}}
The solution of $x_0^-$ is obtained from \lcgii.  The solution
for $y$ is simply found from \xyrelli. Upon quantization these
oscillators obey the standard canonical commutation relations.

The Lagrangian \lcgiii\ in terms of $x$ is similar to the standard
Lagrangian and is expanded in terms of the normal modes in a way
similar to the standard expansion.  The differences in the zero
modes are that $J=-\xi p^+$ must be quantized and the winding term
$ {2\pi w \sigma \tau\over \ell}$ has an unusual form.

The evolution away from $\tau=0$ can be analyzed using either the
$x$ or the $y$ variables. The quantization in terms of $x$ is
standard.  The coordinate $\xi $, which is not subject to
identification on the singular surface $\tau =0$ in $\CO$ is
constrained to be on the lattice $\xi=-J/p^+$.  The evolution of
the nonzero modes is as in standard free string theory.  The
constraint \lzmlzb\ leads to $Jw+N-\tilde N=0$ where $N$ and
$\tilde N$ are the standard number operators.

In terms of $y$ variables we have a single-valued, but time
dependent Hamiltonian \eqn\yahmi{ P_y^-={1\over 4 \pi
\apr}\int_0^\ell d\sigma\
 \left[{(2 \pi \apr \Pi_y )^2\over \tau^2} +
 \tau^2(\partial_\sigma y)^2 \right] = {1 \ov 2 p^+}
\left[{J^2 \over \tau^2} + { w^2 \tau^2 \over \apr^2} \right]
 + {\pi\over \ell} \sum_{n>0} H_n(\tau)
}
where $\Pi_y = {1\over 2\pi \ap} \tau^2 \p_\tau y$ is the
canonical momentum and
 \eqn\oscha{ \eqalign{ H_n(\tau) & =
\left[\lambda_n ( \alpha_n^\dagger \alpha_n + \tilde
\alpha_n^\dagger \tilde \alpha_n) + \rho_n \alpha_n \tilde
\alpha_n + \rho_n^* \alpha_n^\dagger \tilde \alpha_n^\dagger +
\omega _n \right] \cr &\lambda_n=2+\left({\ell\over 2\pi
n\tau}\right)^2\cr &\rho_n=-\left({\ell \over 2\pi n\tau}\right)^2
\left[1+{4\pi in\tau \over \ell}\right]\exp\left(-{4\pi in\tau
 \over \ell}\right) \cr &\omega_n=n\lambda_n}}
The first term on the right hand of \yahmi\ has the interpretation
of the ``rotational kinetic energy'' of the zeromode.  The second
term is the winding energy and the rest is the oscillator
contribution.

Equation \yahmi\ has an interpretation in terms of the Newtonian
expression for the energy a particle with mass $\mu=p^+$ rotating
around an origin at distance $r=\tau$ with angular momentum $J$:
 \eqn\newton{
 H = {J^2\over 2\mu r^2} + V
 }
Note once more the dual role of $x^+$ as both a radial and time
coordinate. The Schrodinger problem with the Hamiltonian \yahmi\
can be solved explicitly \longpaper.

The term ${\pi \over \ell} \sum_{n>0} n\lambda_n(t)$ in the
Hamiltonian \yahmi\ leads to a logarithmically divergent time
dependent subtraction.  It can be interpreted as a logarithmic
subtraction in the definition of $y^-_0$  in terms of $x^-_0$ and
the composite operator $x^2$ \xyrelli.  Therefore, it is absorbed
by a redefinition of the first term $-p^+\partial_\tau y^-_0$ in
the Lagrangian \lcgiii.

Since $\p_y$ is a Killing vector of \ymetric\ we can also
consider the  T-dual background with $g_{yy} \to 1/g_{yy}$ (see
also \refs{\tseytlinup,\smithpolchinski}). Define $\tilde y$ by
the T-duality transformation
 \eqn\tydefa{\eqalign{
 &2\pi \Pi_y={\tau^2\partial_\tau y \over\alpha'}=
 \partial_\sigma \tilde y \cr
 &2\pi \Pi_{\tilde y}={\alpha'\partial_\tau \tilde y \over
 \tau^2}=\partial_\sigma  y \cr}}
where $\tilde y \sim \tilde y+2\pi$. The dual action is
 \eqn\worldsheetyt{S'={1\over 4\pi \alpha'}\int
d \sigma d\tau\ {(\alpha')^2\over \tau^2}\left[
 (\partial_\tau \tilde  y)^2- (\partial_\sigma \tilde y)^2
 \right] + {1 \ov 8 \pi} \int d \sigma d\tau\, \sqrt{\gamma}
\, R^{(2)} \log {\apr \ov (y^+)^2}
}
and the classical solution is
  \eqn\yclasolt{\eqalign{
  \tilde y=&\tilde y_0 - {\xi  \sigma \over \alpha'}+ {2\pi w
  \tau^3\over 3\alpha' \ell}+ \cr
  &p^+\left({\alpha'\over
 2}\right) ^\half \sum_{n\not= 0} \left(1+{2\pi in\tau \over
 \ell}\right) \left({\alpha_n \over n^2}\exp\left[{-
 {2\pi in(\sigma+\tau) \over \ell}}\right] - {\tilde \alpha_n \over
 n^2} \exp\left[{{2\pi in(\sigma-\tau) \over
 \ell}}\right]\right)}}
As expected, the winding number of $\tilde y$, namely $-{\ell \xi
\over 2\pi \alpha'}=-\xi  p^+=J$,  is the eigenvalue of the
parabolic generator $J$, while its momentum is $w=\int d\sigma\
\Pi_{\tilde y}$, which was the winding number of   $y$.

Let us now comment briefly on the evolution of the system when we
include the singular subspace $\CF_{x^+=0}$ or the singular
subspace $\CF_{y^+=0}'$.  We have analyzed in detail the light-cone
quantization of strings moving on both $\CO$ and $\CO'$.
Depending on subtle points regarding the {\it interpretation} of
singular gauge transformations and the geometry of $\CO'$ it
appears that string propagation on $\CO$ differs from that on
$\CO'$. Moreover,  a branelike object, which we call the
``instabrane'' might appear in the theory on $\CO'$. The physical
relevance of the instabrane is a subtle question which will be
addressed in detail in \longpaper.\foot{If instabranes are indeed
not gauge equivalent to Fock space states then a number of
interesting questions arise. For examples: Can they probe the
geometry near the singularity? Are they associated with K-theory
of the crossed-product $C(\IR^{1,2})\sdtimes \Gamma$? }

We remarked above that for an appropriate choice of spin
structure on $\CO - \CF_0 \cong \CO' - \CF_0'$  the orbifold
preserves some spacetime supersymmetries.  It is straightforward
to extend the worldsheet Lagrangian \lcgiii\ to the Green-Schwarz
formalism.  For concreteness consider the model on $\CO\times
\IR^7$.  Before taking the quotient by $\Gamma$ we should add to
\lcgiii\ seven free worldsheet bosons $x^i$ and eight rightmoving
fermions $S^a$ (in the type II theory we also need eight
leftmoving fermions and in the heterotic string also leftmoving
degrees of freedom for the internal degrees of freedom).  It is
easy to see using the symmetries of the problem that after the
action by $\Gamma$ the added fields $x^i$ and $S^a$ remain free
and periodic around the string  \longpaper.\foot{The same
conclusion was also reached by A.\ Tseytlin.}

\newsec{ Covariant Formulation }

\subsec{Quantization and exchange algebra}

The covariant action is
\eqn\covact{
S= {1\over 4\pi \alpha'}\int_{-\infty}^{\infty}
 d\tau \int_0^{2\pi}
 d\sigma \ \eta_{\mu\nu}\left(\p_\tau x^\mu \p_\tau x^\nu
 -\p_\sigma x^\mu \p_\sigma x^\nu\right)
 }
In the twisted  sector the boundary conditions are,
$X(\sigma+2\pi,\tau)= e^{2\pi w \CJ}X(\sigma,\tau) $ where  $w\in
\IZ$ ($\CJ$ was defined in \partrans). The general solution to the
equations of motion in the twisted sector may be expressed in
terms of oscillators
\eqn\oscill{
 \eqalign{
 \hat x_L^\mu:=i \sum_{n\not=
 0} {\alpha_n^\mu \over n} e^{-i n u^+}
\qquad
 &\qquad \hat x_R^\mu:=i \sum_{n\not=
 0} {\tilde \alpha_n^\mu \over n} e^{i n u^-}\cr}
}
where $u^\pm := \sigma \pm \tau$, and the zeromodes,
\eqn\zmode{
X_z(\tau) := \pmatrix{
 x_0^+ + \alpha' p^+ \tau \cr
 x_0 +  \alpha' p \tau \cr
 x_0^- +  \alpha' p^- \tau +
 w^2\bigl( \alpha' p^+ {
 \tau^3\over 6} + x_0^+ {\tau^2\over 2}\bigr)}
}
by introducing a ``spectral flow'' operator:
\eqn\specflow{ \eqalign{ X(\sigma,\tau) = \exp(w \sigma \CJ)
X_z(\tau) &+ \exp(w u^+  \CJ)\hat X_L(u^+) + \exp(w u^- \CJ)\hat
X_R(u^-). \cr} }

The unusual nature of the solution \specflow\ leads to novel
commutation relations on the oscillators. The symplectic form is
standard, $
 \Omega={1 \over 2\pi \alpha' } \int d\sigma\ \delta x^\mu
 \eta_{\mu\nu} \p_\tau \delta x^\nu
$,
but the canonical commutation relations  of the oscillators
become:
\eqn\funnycom{
\eqalign{
 [\alpha_n^{\mu}, \alpha_m^{\nu}]
& = n \delta_{n+m,0}
\left[{1 \ov \eta+ i{ w \ov n} \eta \CJ }\right]^{\mu \nu } \cr
[\tilde \alpha_n^\mu, \tilde \alpha_m^\nu]
& = n \delta_{n+m,0}
\left[{1 \ov \eta - i{w \ov n}\eta \CJ } \right]^{\mu \nu} \cr}
}
while $[\alpha, \tilde \alpha] =0$.

Equations \funnycom\ lead to a curious exchange algebra. One way
to express this is to introduce single-valued fields
$J_\pm(u^\pm):= e^{-w u^\pm \CJ}\p_\pm X$ which satisfy
commutation relations \eqn\pxcom{\eqalign{
 &[j_+^\mu(u^+),j_+^\nu(u^{+'})]= \pi i\alpha' \eta^{\mu\nu}
 \p_+\delta^{(p)}(u^+-u^{+'})+ \pi i \alpha'\delta^{(p)}(u^+-u^{+'})
 F^{\mu\nu}\cr
 &[j_-^\mu(u^-),j_-^\nu(u^{-'})]= -\pi i \alpha'\eta^{\mu\nu}
 \p_-\delta^{(p)}(u^- -u^{-'}) -\pi i \alpha'
 \delta^{(p)}(u^+-u^{+'})  F^{\mu\nu}\cr
 &\qquad F:=w\CJ\eta^{-1}=\pmatrix{
 0&0&0\cr
 0&0&-w\cr
 0&w&0}
 }}
Unlike $\p_\pm X$, the $J_\pm$ are not conformal fields. If we
rotate to Euclidean signature and consider radial evolution in
the complex plane, then \pxcom\ is equivalent to an exchange
algebra in the $w$-twisted sector. For $ \vert z_1\vert > \vert
z_2\vert$ we have
\eqn\exchange{
\p x^{\mu_1}(z_1)    \p x^{\mu_2}(z_2)
=  \p x^{\mu_2}(z_2) \p x^{\mu_1}(z_1)
+ i {w\over z_1 z_2}  \Biggl(
e^{-iw\log z_1\CJ}\CJ e^{i w\log z_2 \CJ}
\eta^{-1}\Biggr)^{\mu_1\mu_2}.
}

The exchange algebra \exchange\  is very similar to the exchange
algebras of chiral vertex operators of RCFT, and indeed the above
analysis applies to {\it any} orbifold by a {\it linear} action
on the spacetime coordinates. Note that the exchange algebra has
the flavor of an Heisenberg algebra, suggestive of non-commuting
coordinates and hence of non-commutative geometry.  Since it is
present only in the twisted sectors with $w\not=0$, and the wound
strings are light only near $x^+=0$, it is reasonable to think
that it reflects a property of the region near $\CF_{x^+=0}$.  In
the context of another time-dependent orbifold, Nekrasov
\NekrasovKF\ has advocated a role of quantum groups.  He
discusses a two-dimensional model, and considers D0 branes, while
we are working in the chiral sector of the closed string.  The two
results are different, but they are similar in spirit. Possible
noncommutativity in light-cone coordinates was also discussed in
\KoganNN.

\subsec{The Physical state conditions}

Although the worldsheet energy-momentum tensor has
a standard form in $X$-coordinates:
\eqn\strt{
 T_{++} = {1\over \alpha'} \p_+ X^{tr} \eta \p_+ X = \sum_{n\in Z}
 L_n e^{in(\tau+\sigma)}
 }
we have a  nonstandard  realization of Virasoro operators since
$\CJ$ is not diagonalizable, making the solution of the physical
state conditions  nontrivial. Nevertheless, a DDF-operator
construction shows the BRST cohomology is nontrivial in the
twisted sectors. More explicitly, it is easy to construct
physical states in the winiding sector with zero modes only
\eqn\orthcompl{ \sqrt{p^+\over i x^+_0}
 \exp\Biggl[-ip^+ x^-_0 - i{\m^2\over 2p^+}
 x^+_0 + i{p^+ \over 2 x^+_0}(x_0+{J \over p^+})^2 -
  i{w^2 (x^+_0)^3\over 6(\alpha')^2p^+} \biggr]
 }
%%%
where $\m^2 := m^2 + \vec{p}_{\perp }^2$, while
$\vec p_\perp$ is the momentum in the transverse directions
and $m^2 = - {4 \ov \apr}$.  One
then can show that DDF-like operators \eqn\defddf{A_n = \oint {dz
\ov 2 \pi} \, \left({2 \ov \apr}\right)^{\ha} \, \biggl[ \p_z x +
i w \log z \p_z x^+ +{w\over znk_0}\biggr] \, e^{i n k_0 x^+}
(z), \qquad n = \pm 1, \pm 2 \cdots
 }
with  $ k_0 = {2 \ov \apr p^+} $, are well-defined and survive
the orbifold projection. Together with their rightmoving
counterparts $\tilde A_n$, they act on \orthcompl\ to generate
physical states with a full tower of positive signature Fock
space states, even in the twisted sectors. This will be confirmed
by the computation of the trace in covariant quantization.

\subsec{Torus partition function}

Let us now consider the path integral on a torus with worldsheet
metric
 \eqn\torusm{ g= dz_+ dz_-= (d\si + \tau_+ d \sii)(d\si +
 \tau_- d\sii)}
where $\si,\sii$ have period 1, while $\tau_\pm\in \IR$ for
Lorentzian signature, and $\tau_+ = (\tau_-)^*$ for Euclidean
signature tori.  Until further notice we will use a Lorentzian
torus.

In the orbifold theory we sum over winding sectors $(w_a,w_b)$
around the $a,b$-cycles. In these sectors we write the field as
\eqn\torone{
X(\si,\sii) = \exp\bigl[ 2\pi(\si w_a + \sii w_b) \CJ \bigr]
\sum_{n_a, n_b \in \IZ}
 X_{n_a, n_b} e^{2\pi i (n_a\si + n_b \sii)}.
}

Because the matrix $\CJ$ is strictly lower triangular the
contribution of the quantum fluctuations is {\it independent} of
$w_a,w_b$, and one finds that the one-loop partition function for
the string theory is:
%%%
%
 \eqn\toramp{ \eqalign{ & Z =
  \int_{\CO} ~~ \! \! \!
 {d^3 x \over (2\pi\sqrt{\ap})^3} \! \! \! \sum_{w_a,w_b\in \IZ}
 \!\! \! \exp\Biggl[- i \pi
 {(x^+)^2\over \ap } {(w_b + w_a \tau_+) (w_b+w_a \tau_-) \over
 \tau_2} \Biggr]  {-iZ^{\rm ghost}Z^{\perp }(w_a,w_b) \over (-i
 \tau_2)^{3/2} (\eta(\tau_+)\eta(-\tau_-))^{3}}
 \cr}
}
with $\tau_2=(\tau_+-\tau_-)/2$.  This formula can also be
derived by explicit evaluation of the trace
\eqn\offshelltr{
Z^{\rm matter} =
{\Tr}_{\CH}  e^{2\pi i \tau_+ (L_0-c/24)} e^{-2\pi i
\tau_-( \tilde L_0-c/24)}
}
where $\CH$ is the CFT state space of the orbifold. In this
second derivation it is important to take $\tau_\pm$ real since
$L_0$ is not bounded below.

Of course, \toramp\ bears a striking resemblance to the partition
function for the partition function for a Gaussian field with
target space a circle of radius $x^+$. Once again, the dual role
of $x^+$ as both a radial and a time variable leads to some
interesting physics.

Note that we have left the integration over the zeromode
$x^\mu=X_{00}^\mu$ of the field $X$ undone in \toramp.\foot{Of
course, we could perform the $x^+$ integral. In this case we
obtain a nonholomorphic Eisenstein series.} This is important for
the correct conceptual interpretation of the amplitude: We have a
spacetime-dependent contribution to the cosmological constant
%%%
%
\eqn\lambdax{ \eqalign{ & \Lambda(x^\mu) =  {i\over (2\pi\sqrt{\ap})^3}
\sum_{w_a,w_b\in \IZ}
\int_{{\cal F}}
{d\tau_+ \wedge d\tau_- \over (\tau_2)^2}
 \exp\Biggl[-i \pi {(x^+)^2\over \ap }
{(w_b + w_a \tau_+) (w_b+w_a \tau_-) \over \tau_2} \Biggr] \cr
& \qquad \qquad \qquad \qquad \times \, \,
 {Z^{\perp} (w_a,w_b)\over (-i
 \tau_2)^{1/2} (\eta(\tau_+)\eta(-\tau_-))} \ \  . \cr} }
At this point, we will take the torus to have Euclidean
signature. Some of the issues that arise when attempting to make
sense of the Lorentzian signature formulae were discussed in
\MooreZC\ and will be further addressed in \longpaper.

The expression \lambdax\ exhibits a curious divergence as $x^+\to
0$. Suppose, for simplicity, that $\Gamma$ does not act on
$\CC^{\perp}$. Then we may also write:
 \eqn\comxp{\Lambda(x^+) = -  { 1\over (2\pi)^3 \ap \vert
 x^+\vert }\int_{{\cal F}} {d^2 \tau  \over (\tau_2)^2}
 {Z^{\perp}\over \vert \eta(\tau)\vert^2} \sum_{w_a,\hat w_b\in
 \IZ} q^{\half \ap p_L^2} \bar q^{\half \ap p_R^2}}
with $ p_L = {1\over \sqrt{2}} ({\hat w_b\over x^+} + w_a
{x^+\over \ap}) $ and $ p_R = {1\over \sqrt{2}} ({\hat w_b\over
x^+} - w_a {x^+\over \ap})$. The modes with $\hat w_b=0,
w_a\not=0$  are {\it light winding modes} for $x^+\to 0$. This
shows there is a {\it divergence} as $x^+\to 0$:
\eqn\divergence{ \Lambda(x^+)\sim {1\over (x^+)^2} \times
\int_{{\cal F}} {d^2 \tau  \over (\tau_2)^2}  {Z^{\rm tr}\over
 \vert \eta(\tau)\vert^2} }
This divergence has a simple interpretation as  a volume
divergence in the $T$-dual coordinate $\tilde y$ of radius
$1/x^+$. This suggests that the light winding strings ``open up''
the singularity from a cone to a trumpet. Of course, we cannot
conclude this is the correct qualitative physics until we have
carefully considered one-loop interactions and issues of
backreaction.

The above formulae apply to the superstring in the NSR formalism.
In this case $Z^\perp$ includes the contributions of the NSR
fermions. When we choose the supersymmetric spin structure on the
space time fermions the phases in the sum over the spin
structures and the partition functions themselves are independent
of $(w_a,w_b)$, and hence \divergence\ multiplies zero. If we
choose the other spin structure in spacetime, the situation is
similar to that in \rohm.  Then there are tachyons for $\vert
x^+\vert$ smaller than the string scale and a negative
cosmological constant is generated.

\newsec{Interactions}

\subsec{Tree level interactions}

We have studied tree level interactions in $\phi^3$ field theory
on the orbifold, in string theory in light-cone gauge, and in the
covariant formulation.  Here we briefly summarize some results
about tree level amplitudes deferring more details to \longpaper.
The main result is that for generic momenta the amplitudes are
finite, but when the intermediate particles have $p^+=0$ the
amplitudes can diverge.  These divergences are associated with the
singularity at $x^+=0$.

At large $|x^+|$, the geodesic distances between the image points
go to infinity; i.e.\ the space effectively opens up. Thus we
expect to be able to prepare our ``in'' and ``out'' states in the
far past and far future, and to consider the S-matrix elements
similar to those in $\IR^{1,2}$. The standard S-matrix elements
in $\IR^{1,2}$ are expressed in the plane wave basis \phidef.
However, the plane waves are not invariant under the orbifold
action and the invariant functions constructed from them by
summing over the images are not convenient to work with.  A
better basis is the $J$ basis $\psi_{p^+,J}$ of \wavefun, in
terms of which the orbifold projection simply corresponds to
taking $J$ to be integral. This motivates us to consider the
S-matrix elements in flat $\IR^{1,2}$ in the $J$ basis. This is
somewhat unusual since going to the $J$-basis introduces time
dependence which arises because $J$ and $p^-$ do not commute.
More explicitly, the wave function $\psi_{p^+,J}$ of a free
particle is forced to be localized at $\xi= -J/p^+$ at $x^+ =0$,
which breaks the translational invariance of $x^+$. In other
words, the on-shell ``in'' and ``out'' states are prepared in a
way that they are aimed at points $\xi_i = - J_i/p^+$ ($i$ labels
particles) so that at $x^+ =0$, their wavepackets  are completely
localized at the respective points. In flat $\IR^{1,2}$, this
corresponds to asking time-dependent questions in a time
independent background.  S-matrix elements in the $J$-basis are
observables in $\IR^{1,2}$  which exhibit some features of time
dependent amplitudes.

Since $\psi_{p^+,J}$ is obtained from the plane waves by a Fourier
transform \wavefun, the amplitudes in the $J$ basis can be
obtained from those in the standard momentum basis by a Fourier
transform. Once these tree level S-matrix elements are computed
in flat space, it is straightforward to compute them for the
orbifold simply by restricting $J_i$ to be integers.

In the following we give the results for the three-point and
four-point tree-level amplitudes for untwisted tachyons on the
orbifold. More details will appear in \longpaper.

{}From \wavefun\ the vertex operator for the tachyon in the $J$-basis
can be written as
 \eqn\verJ{ V_{p^+_i,J_i} (\sigma,\tau) =  {1 \over \sqrt{2 \pi
 p_i^+}}\ \int_{-\infty}^\infty dp \,
 e^{ip \xi_i} \, e^{i \vec{p_i} \cdot \vec{X}(\sigma,\tau)},
 \qquad \xi_i=-{J_i \ov p^+_i}}
where $i$ labels external particles, the factor of $1\over
\sqrt{p_i^+}$ is introduced for the natural normalization of
states in the light-cone frame, and
 \eqn\veropa{e^{i \vec{p}_i \cdot \vec{X}(\sigma, \tau)}
 = \exp \left[-i p^+_i x^- - i p^-_i x^+ + i p_i x +
 i \vec{p}_{\perp i} \cdot \vec{x}_{\perp} (\sigma, \tau) \right]}
($\vec p_{\perp i}$ and $\vec x_{\perp i}$ denote vectors in the
transverse dimensions) is the standard on-shell tachyon vertex
operator with
 \eqn\momdefa{p^-_i = {p^2_i + \m_i^2 \ov 2
 p^+_i}, \qquad \m_i^2= m^2 + \vec{p}_{\perp i}^2,\qquad
   m^2 = - {4 \ov \apr} \ . }

\bigskip

\centerline{\it Three point function}

\bigskip

The  $1\to 2 + 3$ tachyon amplitude in the $J$-basis in the
covering space is \longpaper\
 \eqn\mampt{ A_3 = {8 \pi i g_s \ov \apr} (2
 \pi)^{25} \, \delta (p_1^+ - p_2^+ - p_3^+) \, \delta (\vec
p_{\perp 1} - \vec p_{\perp 2} -
 \vec p_{\perp 3}) \, \delta(J_1-J_2 - J_3)  \, w_3 (J_i,p_i^+) }
with
  \eqn\deflwth{\eqalign{
 w_3(p^+_i; J_i)  = & \int^{\infty}_{-\infty} d x^+  \,
 {1 \ov (- i x^+)^{1\ov 2}} \, e^{-{i \ov 2} \alpha (p^+_i, \m_i)
 x^+ - {i \ov 2 x^+}\mu_{23}(\xi_2-\xi_3)^2}\cr
 =& \cases{  2  \sqrt{2  \pi \ov \al}  \cos \left(\sqrt{\al
 \mu_{23}}(\xi_2-\xi_3)\right), & $  \al > 0 $ \cr \cr 0, &
 $\al < 0$ \ .} }}
In \deflwth\ we have defined
 \eqn\defab{ \eqalign{
  \alpha (p^+_i,\m_i)& = {\m_1^2 \ov p_1^+} -{\m_2^2 \ov p_2^+}
  - {\m_3^2  \ov  p_3^+}  \cr
   \mu_{23}&= {p_2^+p_3^+\over p_2^+ + p_3^+}}}
Going to the orbifold,  we take $J_i$ to be integers and replace
the Dirac delta function for $J_i$ by $\delta_{J_1,J_2+J_3}$ as a
result of dividing the amplitude by the volume of the orbifold
group.

It is worth noting that the amplitude is not an analytic function
of $\alpha$.  Such lack of analyticity in the ``momenta'' is
common in problems with a noncompact inhomogeneous target space
such as the $c=1$ system
\refs{\PolchinskiMF\PolchinskiUQ\DiFrancescoSS\DiFrancescoUD-\MooreGB}.
The novelty here is that the inhomogeneous direction is not
spacelike but lightlike.

These amplitudes have a simple physical interpretation in the
light-cone description of the theory.  We have already mentioned
that in the light-cone description we interpret the system as
nonrelativistic particles with time, mass and potential energy
 \eqn\notation{\tau=x^+, \quad \mu_i=p^+_i, \quad
 V_i=\m_i^2/2\mu_i}
In this notation the functions $\psi_{p^+,J}=\sqrt{\mu\over i
\tau}e^{ -i\mu x^- -iV \tau+ i{ \mu \over 2 \tau} (x-\xi)^2}$ are
the nonrelativistic propagator from a position $x$ at time $\tau$
to the position $\xi=-{J\over p^+}$ at time $\tau=0$ times the
trivial factor $e^{-i\mu x^--iV \tau}$.  This is another way to
understand the focusing at $x=\xi$ at $x^+=0$.

In terms of the notation \notation\ the decay process is
described as a particle with mass $\mu_1$ decaying to two
particles with with masses $\mu_2$ and $\mu_3$.  $p^+$
conservation, which arises from translational invariance of
$x^-$, is being interpreted as conservation of mass
 \eqn\masscon{\mu_1=\mu_2+\mu_3 \ .}
The center of mass coordinate of the decay products is
${\mu_2x_2+\mu_3x_3 \over \mu_2+\mu_3}$.  $J$ conservation
guarantees that its value at $x^+=0$ is the same as the value of
$x_1$ at $x^+=0$
 \eqn\centerofmass{\xi_1={\mu_2\xi_2+\mu_3\xi_3 \over
 \mu_2+\mu_3}}
The remaining dynamics is best described in terms of the relative
coordinate and the reduced mass
 \eqn\relcred{x_{23}=x_2-x_3, \qquad
 \mu_{23}={\mu_2\mu_3 \over \mu_2+\mu_3} }
The parameter $\al$ defined in \defab\ takes the form
 \eqn\defabm{ \al = 2(V_1-V_2 - V_3) }
and is interpreted as the difference in potential energy between
particle $1$ and particles $2$ and $3$. An obvious necessary
kinematical condition for the decay is $\alpha=2(V_1-V_2-V_3)>0$,
as is the case in \deflwth.  This also explains why the amplitude
is nonanalytic as a function of $\alpha$.

The decay amplitude \deflwth, can be rewritten as
 \eqn\defsmwa{w_3(\mu_i,\xi_i)  = \int^{\infty}_{-
 \infty} d x^+ \, {1\over \sqrt{-i x^+}} \, \exp \left[
 i(V_2+V_3-V_1) x^+ - i{\mu_{23}(\xi_2-\xi_3)^2 \over 2 x^+}
 \right] \ . }
We see an integral over the time of the interaction $x^+$.  The
integrand has two phases.  The first term, as discussed above, is
associated with the difference in potential energy.  The second
term, is the free propagation of the relative coordinate of the
decay products from its value $x_{23}=\xi_2-\xi_3$ at $x^+=0$ to
its value $x_{23}=0$ at the time of the interaction $x^+$.

\bigskip

\centerline{\it Four point function}

\bigskip

The four point function probes the structure of the interacting
theory in more detail than the three point function.  As a
particle approaches the singularity it is blue shifted, its
energy becomes large, and its coupling to the graviton becomes
large. Therefore, one expects large back reaction of the geometry
which could appear as a divergence in the four point function.
The same argument can be used for the correlation functions of
\verJ\ in the covering space because the focusing at $\xi_i$
leads to large energy density and potentially large back reaction.
%%%
A signal of such a problem would be the
failure of the convergence of the Fourier transform of the
Virasoro-Shapiro amplitude. We therefore study this transform in
some detail.

The Fourier transform of the Virasoro-Shapiro amplitude can be
written as
 \eqn\ndire{
 A_4 = {8 (2 \pi)^3 i g_s^2\over \alpha'}
  \, \int \left(\prod_{i=1}^4 {d p_i \ov
 \sqrt{2 \pi p_i^+}}\right)\, \delta (p_1 +p_2 -p_3-p_4) \,
 e^{i F} \, \delta(E) \, A(s,t)  \,
 }
where we suppressed a factor of $(2\pi)^{24}\delta(p_1^++p_2^+-
p_3^+-p_4^+)\delta(\vec p_{\perp 1} + \vec p_{\perp 2} - \vec
p_{\perp 3}- \vec p_{\perp 4})$ and
 \eqn\defanc{ \eqalign{
 A(L_s,L_t,L_u) =& \pi {\Ga \left(-{\apr \ov 4} L_s\right)
 \Ga\left(-{\apr \ov 4} L_t\right) \Ga \left(-{\apr \ov 4}
 L_u\right)
 \ov \Ga  \left(1+{\apr \ov 4} L_s\right)   \Ga \left(1+ {\apr \ov
 4} L_t\right) \Ga \left(1+ {\apr \ov 4} L_u\right)}\cr
 =& - \left({\Ga \left(-{\apr \ov 4} L_t\right)  \Ga
 \left(- {\apr \ov 4} L_u\right) \ov \Ga  \left(1+{\apr \ov 4}
 L_s\right)}\right)^2 \, {\sin \left({\apr \pi \ov 4} L_t\right) \,
 \sin \left({\apr \pi \ov 4} L_u \right) \ov \sin \left({\apr \pi
 \ov 4} L_s \right)  } }}
and
 \eqn\para{\eqalign{
 E & = p_1^- + p_2^- - p_3^- - p_4^- ={p_1^2 + \m_1^2 \ov2 p_1^+}
 + {p_2^2 + \m_2^2 \ov 2 p_2^+} - {p_3^2 + \m_3^2 \ov2 p_3^+}
 - {p_4^2 + \m_4^2 \ov2 p_4^+} \cr
 F & = p_1 \xi_1 + p_2 \xi_2 - p_3 \xi_3 - p_4 \xi_4  \cr
 L_s & =  s - m^2 + i \ep = 2( p_1^+ +  p_2^+)( p_1^- +  p_2^-) -
 ( p_1 +  p_2)^2  - \m_s^2 + i \ep \cr
 L_t & =  t - m^2 + i\ep = 2( p_1^+ -  p_3^+)( p_1^- -  p_3^-) -
 ( p_1 -  p_3)^2 - \m_t^2 + i\ep\cr
 L_u & =   u - m^2 + i \ep =2( p_1^+ -  p_4^+)( p_1^- -  p_4^-) -
 ( p_1 -  p_4)^2 - \m_u^2 + i \ep\cr
 \m_s^2& =m^2+(\vec p_{\perp 1}+\vec p_{\perp 2})^2, \quad
 \m_t^2 =m^2+(\vec p_{\perp 1}-\vec p_{\perp 3})^2, \quad
 \m_u^2=m^2+(\vec p_{\perp 2}-\vec p_{\perp 3})^2 }}

The momentum integrals can be reduced to a single integral
expression
 \eqn\vadell{
  A_4 ={8 (2 \pi)^2 i g_s^2\over \alpha'}
  \, \delta(J_1+J_2-J_3-J_4)
  \int_{-\infty}^{\infty}  {d q \ov |q|} \,
  \exp\left[{i \ov 2} \left(q \xi_{-} +  {\alpha \xi_+ \ov
  q}\right) \right] \; A(L_s,L_t,L_u)  }
with
 \eqn\panewv{\eqalign{
 L_s & = (p_1^+ + p_2^+)\left(q_+^2 +
  {\m_1^2 \ov p_1^+} +{\m_2^2  \ov  p_2^+}  \right) - \m_s^2
 + i \ep \cr
 L_t & =  (p_3^+ - p_1^+)\left( {\m_3^2  \ov  p_3^+} -{\m_1^2 \ov
 p_1^+}\right) - \m_t^2 - \mu_{12}
 \left(\sqrt{p_3^+\over p_1^+} q_+ -\sqrt{p_4^+\over p_2^+ }
 q_-\right)^2 + i \ep \cr
 L_u    & =  (p_4^+ - p_1^+) \left({\m_4^2  \ov  p_4^+}
 -{\m_1^2 \ov p_1^+} \right)- \m_u^2 -
 \mu_{12}\left(\sqrt{ p_4^+\over p_1^+ } q_+ +\sqrt{p_3^+ \over
 p_2^+ } q_- \right)^2 + i \ep \cr
 \xi_\pm  & =\sqrt{\mu_{12}}(\xi_1-\xi_2)\pm \sqrt{\mu_{34}}
 (\xi_3-\xi_4)\cr
 q_\pm&=\half\left(q \pm {\alpha\over q}\right)\cr
  \mu_{12} & = {p_1^+ p_2^+ \ov p_1^+ + p_2^+}, \qquad
   \mu_{34} = {p_3^+ p_4^+ \ov p_3^+ + p_4^+}\cr
 \alpha & =  {\m_3^2 \ov p_3^+} + {\m_4^2  \ov  p_4^+}
 - {\m_1^2 \ov p_1^+} -{\m_2^2  \ov  p_2^+}\cr
 }}
This is an integral over various values of $s$ with corresponding
values of $t$ and $u$. The integral passes through the poles in
the s-channel.  These do not lead to divergences because of the
$i\epsilon$.  Instead, these poles make the amplitude nonanalytic
as a function of the external momenta $\vec p_i$.  This
nonanalyticity is similar to the nonanalyticity we have already
observed as a function of $\alpha$ in the three point function
and its origin is similar.  It is associated with large $|x^+|$.

Let us analyze the amplitude in more detail and consider first the
situation of generic $p_i^+$. Possible divergences and
nonanalyticity in $\xi_i$ can arise only from the behavior of the
integral \vadell\ at $q=0,\pm\infty$, which correspond to the
interesting region $x^+\approx 0$. For generic $p_i^+$ each of
these limits corresponds to the hard scattering limit of large
$s,t,u$ with fixed ratios. In this limit $A(L_s,L_t,L_u)$ decays
rapidly and the integral \vadell\ converges for all $\xi_i$.  The
dependence on $\xi_i$ is analytic.

We would like to contrast this result with the corresponding
behavior in field theory.  Because the large momentum dependence
of the field theory amplitude is power like, the Fourier
transforms do not share the nice analytic structure in $\xi$ that
the string amplitudes enjoy.  We conclude that the better UV
behavior of string theory improves the analytic structure of the
amplitudes as a function of $\xi_i$.

For nongenric $p_i^+$ the situation is different. Consider the
kinematical configuration with vanishing
$p_t^+=p_3^+-p_1^+=p_2^+-p_4^+$.  Then the large $q$ behavior of
\vadell\ is in the Regge region and $A(L_s,L_t,L_u) \sim
q^{-\alpha'\m_t^2} = q^{4-\alpha' \vec p_{\perp t} ^2}$ with
$\vec p_{\perp t}= \vec p_{\perp 3} -\vec p_{\perp 1 }$.  When
the exchanged particle has vanishing $J_t=J_3-J_1=0$, i.e.\
$\xi_1=\xi_3$, and $\xi_2=\xi_4$, the integral over $q$
diverges\foot{When $\xi_1 \neq \xi_3$, the integral converges.
Physically, the convergence for $\xi_1 \neq \xi_3$ arises since
the external particles are not focused at the same point.} for
$\alpha' \vec p_{\perp t}^2 <4$.  To see the divergence in more
detail consider the scattering with $J_t=0$ and
$p_t^+=p_3^+-p_1^+ \to 0$.  Then the amplitude behaves as $A_4
\sim (p_t^+)^{-4+\alpha'\vec p_{\perp t} ^2}$, up to
logarithmic corrections. We interpret this
behavior which arises from the Regge region as due to the
exchange in the t-channel of particles with ``spin'' $ \alpha(t)=
2-\half\alpha'\vec p_{\perp t}^2$.  For $\vec p_{\perp t}=0$ they
include the graviton. By exchanging $3\leftrightarrow 4$ we find
a similar situation in the u-channel.

We conclude that even though the amplitude for fixed generic
momenta is finite, the total cross section diverges due to
singularities associated with particles with $p^+=0$ in the t and
the u channel.  A more detailed analysis of the amplitude shows
that these particles are exchanged at $x^+\approx 0$, and
therefore can be interpreted as associated with the singularity
and the large energy density there.  We do not fully understand
the implication of this phenomenon and it might be a signal of
the breakdown of perturbation theory.

It is important to point out that the amplitude we have just
derived cannot be obtained by a straightforward application of the
formalism based on Euclidean worldsheets.  Since the target space
has Lorentzian signature, we expect the worldsheet to have
Lorentzian signature.  Furthermore, a Fourier transform similar to
\ndire\ of the standard density on the Euclidean signature moduli
space need not converge. On the other hand, an expression for the
amplitude with a Lorentzian worldsheet appears to exist, and to
have a nonrelativistic interpretation similar to that of the
three point function.

\subsec{Tadpoles and backreaction}

We have remarked above on the question of backreaction of the
geometry to incoming particles. Another important question is
whether our background is stable against quantum fluctuations.
Since it does not have closed timelike curves, we do not expect
Hawking's chronology conjecture \HawkingNK\ to demand the
existence of new singularities. However, the presence of closed
null curves at $x^+=0$ is potentially dangerous.  In standard
orbifolds the expectation values of composite operators like the
stress tensor diverge at fixed points.  In Lorentzian orbifolds
such divergences occur along closed null curves (see e.g.\
\refs{\MaldacenaKR,\HiscockJQ}, and references therein). For
example, the expectation value of the stress tensor of a free
scalar field in the similar problem of a BTZ black hole was
calculated in \refs{\ShiraishiHF\SteifZV-\LifschytzEB}, and for
fermions in \KimMI\ (for a review see \CarlipQV).  Such an
expectation value acts as a source for the gravitational field
and destabilizes the background.

We have already seen that in the background with the
nonsupersymmetric spin structure a cosmological constant is being
generated.  It is therefore clear that this background is
unstable.  A cosmological constant is not generated at one loop
in the supersymmetric model, and therefore it makes sense to
examine this background in more detail.  More specifically we
consider $\IR^7\times \CO$. Using the symmetries of our problem
(supersymmetry and the two Killing vectors $\partial_y$ and
$\partial_{y^-}$) and the results of \mwaver\ it is
straightforward to show that the most general metric and dilaton
field are
 \eqn\mostgen{\eqalign{
 &(ds)^2=-2dy^+dy^- + (f(y^+))^2 (dy)^2+ (dx^i)^2\cr
 &D=D(y^+)}}

At tree level $f(y^+)=y^+$ and $D$ is a constant. This raises the
question of whether they have radiative corrections. We have
examined the one-loop tadpoles for on shell physical particles,
and found them to vanish. The only possible tadpoles come from the
graviton with polarization along the orbifold or the dilaton. In
the tadpole computation we examined the torus expectation values
of these vertex operators with  arbitrary $p^-$, since the
background depends on $x^+$. In the covariant formulation the
expectation values turn out to vanish because of the sum over the
spin structures. In the light-cone Green-Schwarz formalism they
vanish because of the presence of fermion zero modes. Details
(and subtleties) of these computations will be discussed in
\longpaper.

We conclude that up to one loop order the zero and one point
functions vanish and our background is stable.  However, the
subtleties with intermediate particles with $p^+=0$ we
encountered in the tree level four point function prevent us from
concluding that the background is absolutely stable.  Such
particles can appear in the two point function at one loop or the
zero point function at two loops and could potentially
destabilize our orbifold.

\newsec{Conclusions and Future Directions}

We have studied the parabolic orbifold following the
standard rules of perturbative string theory. We have
found that the parabolic orbifold makes suprisingly
good sense, despite the preponderance of potentially
pathological pitfalls.  One cloud on the horizon is
the divergence of the 4 point function in special
kinematic configurations. This needs to be understood
much more thoroughly.

The issue of backreaction is currently under study, and might
prove to be a serious problem with future development of this
example. One way backreaction could ruin the orbifold is through
the coupling of gravity to the large energy momentum of particles
which are blue shifted near the singularity.  Indeed, we found
that although the tree level four point function is finite for
generic momenta, it diverges when the exchanged particle in the t
or u channel has vanishing $p^+$.  This could signal the
breakdown of perturbation theory.   An interesting possibility,
which has been suggested by numerous physicists, is that every
scattering process might create a black hole.
%
% Perhaps this is
%what the behavior we found at $p^+_t=0$ is trying to tell us.
%

In the spirit of black hole complementarity \SusskindIF\ it might
be that the system admits two complementary descriptions.  One of
them, for an observer at $x^+<0$ is in terms of a Universe which
ends at the singularity.  The other description includes both
sides of the space on the two sides of the singularity.

We have found that there are tantalizing hints of a role of
noncommutative geometry in resolving the $x^+=0$ singularity, and
in making sense of string theory on the non-Hausdorff quotient
space. Perhaps related to this is the  nontrivial exchange
algebra for the coordinate currents $\p x^\mu$ in the twisted
sectors, a result which is possibly related in turn to that of
\NekrasovKF.

We believe there are many other interesting examples of
noncompact orbifolds associated with noncompact discrete groups.
The flavor of these models is rather different from that of the
compact orbifolds by finite groups which have been much studied in
the context of Calabi-Yau compactification. This new territory
might hold some exciting new surprises.

Note added: After this paper was submitted it was suggested by
several people and was shown more decisively in
\refs{\lawrence\hp-\lmst} that the singularity in the four point
function indeed reflects a large backreaction of the geometry to
incoming particles.  This backreaction renders perturbation
theory invalid.  However, the techniques presented here are
useful in analyzing closely related models which are not singular.

\bigskip
\centerline{\bf Acknowledgements}

We thank B. Acharya and  A. Tseytlin for participating in the
early stages of this project and discussions. We also thank
T.~Banks, S.~Frolov, J.~Harvey, G.~Horowitz, D.~Kutasov,
J.~Maldacena, E.~Martinec, A.~Strominger, C.~Thorn and E.~Witten
for very helpful discussions. HL and GM were supported in part by
DOE grant \#DE-FG02-96ER40949 to Rutgers. NS was supported in part
by DOE grant \#DE-FG02-90ER40542 to IAS. GM would like to thank
the Isaac Newton Institute for hospitality during the completion
of this manuscript.

\listrefs

\bye